# δ-phosphorene: a two dimensional material with high negative Poisson's ratio

*Haidi Wang,[a] Xingxing Li,[ab] Pai Li[ab] and Jinlong Yang[\*ab]*

[a]Hefei National Laboratory of Physical Science at the Microscale, University of Science and Technology of China, Hefei, Anhui 230026, China

[b]Synergetic Innovation Center of Quantum Information & Quantum Physics, University of Science and Technology of China, Hefei, Anhui 230026, China

E-mail: jlyang@ustc.edu.cn

**ABSTRACT**

As a basic mechanical parameter, Poisson's ratio (ν) measures the mechanical responses of solids against external loads. In rare cases, materials have a negative Poisson's ratio (NPR), and present an interesting auxetic effect. That is, when a material is stretched in one direction, it will expand in the perpendicular direction. To design modern nanoscale electromechanical devices with special functions, two dimensional (2D) auxetic materials are highly desirable. In this work, based on first principles calculations, we rediscover the previously proposed δ-phosphorene (δ-P) nanosheets [Jie Guan et al., *Phys. Rev. Lett*. 2014, 113, 046804] are good auxetic materials with a high NPR. The results show that the Young's modulus and Poisson's ratio of δ-P are all anisotropic. The NPR value along the grooved direction is up to -0.267, which is much higher than the recently reported 2D auxetic materials. The auxetic effect of δ-P originated from its puckered structure is robust and insensitive to the number of layers due to weak interlayer interactions. Moreover, δ-P possesses good flexibility because of its relatively small Young's modulus and high critical crack strain. If δ-P can be synthesized, these extraordinary properties would endow it great potential in designing low dimensional electromechanical devices.

**INTRODUCTION**

Poisson's ratio and Young's modulus are of great importance to evaluate a material's mechanical strength and stability[1–5], and also crucial to design high performance electromechanical devices. Poisson's ratio ($\nu$)[6–8], defined by the ratio of the strain in the transverse direction to that of the longitudinal direction, measures the fundamental mechanical responses of solids against external loads. In general, when a compressive (tensile) stress is acted in one direction, materials tend to expand (contract) in the perpendicular direction. For these materials, Poisson's ratio has a positive value. The Poisson's ratio of common solid state crystals usually fall in the range of $0<\nu<0.5$, while gases and cork[9,10] have $\nu\approx0$. In rare cases, a negative Poisson's ratio (NPR)[7] is attainable, which shows a strong correlation with atomic packing density, atomic connectivity[8] and structural phase transition.[11,12] These materials are called auxetic materials, and typically have enhanced toughness and shear resistance, as well as significant sound and vibration absorption. Auxetic materials have been exploited in fields such as medicine, fasteners, tougher composites, national security and defense.[13]

So far, the search for NPR materials has mainly been focused on bulk and engineered auxetic structures. In 1987, foams with negative Poisson's ratios were firstly produced by Lakes[14] from conventional low density open-cell polymer foams by causing the ribs of each cell to permanently protrude inward. For this auxetic material, the NPR arises from its re-entrant structure. Later, auxetic phenomena have also been observed in crystalline $SiO_2$ and other cubic materials.[15–18]

To fabricate electromechanical devices at nanoscale, low dimensional auxetic materials are very desirable. In recent years, there have been increasing interests for exploring the possibility of auxetic phenomenon in low dimensional systems.[1,13,19–22] For example, theoretical calculation has forecasted that black phosphorene (BP)[23] possesses an intrinsic auxetic effect resulting from its puckered configuration,[1] although the NPR value is comparatively small (-0.027). In addition, auxetic phenomena have also been predicted in borophene[20], penta-graphene[24], penta-boron nitride[22], $Be_5C_2$[25] and SnSe[19]. However, the NPR values of all these 2D materials are relatively small, hindering their further applications. In order to design high performance nanoscale electromechanical devices, 2D materials with high NPRs are urgently needed.

In this work, based on first-principles calculations, we further study the orientation-dependent Young's modulus, Poisson's ratio, quantum size effect and electronic structure of previously proposed $\delta$-P[26] (See Fig. 1). Interestingly, we discover that $\delta$-P is a superior auxetic material with a NPR value as large as -0.267. Like its allotrope BP[3], $\delta$-P has a puckered structure with grooves formed by two layers of phosphorus atoms. Previous study[26] demonstrates that $\delta$-P is stable up to 1000 K. Besides, δ-P has good electronic properties. For example, the band gap of δ-P is 0.66 eV (HSE06 level)[27], which indicates that δ-P has a potential application for designing the nano electro-optical devices with infrared light absorption.

**COMPUTATIONAL METHODS**

The first principles calculations were carried out based on the Kohn-Sham density functional theory[28] (KS-DFT) as implemented in the Vienna ab initio simulation package (VASP).[29] The generalized gradient approximation as parameterized by Perdew, Burke and Ernzerhof[30] (PBE) for exchange-correlation functions was used. Electronic wave functions were expanded in a plane wave basis set with the kinetic energy cutoff of 450 eV. The convergence criterion for the energy in electronic SCF iterations and the force in ionic step iterations were set to be $1.0\times10^{-6}$ eV and $5.0\times10^{-3}$ eV/Å, respectively. The non-periodic direction was set along z direction and at least 15 Å vacuum slab was added to eliminate the interaction between the δ-P and its replicas resulted from the periodic boundary condition. The reciprocal space was sampled with a k-grid density of $0.04\times2\pi$ Å$^{-1}$ for the ionic iterations and $0.02\times2\pi$ Å$^{-1}$ for electronic SCF iterations using the Monkhorst-Pack method. Besides, van der Waals[31–33] (vdW) correction proposed by Grimme (DFT-D2) was chosen due to its good description of long-range vdW interactions. All the mechanical properties were calculated by PBE functional with DFT-D2 correction unless otherwise stated. Phonon properties are calculated using finite displacement method implemented in Phonopy[34]. A (3 × 3) supercell was constructed to calculate the atomic forces by using VASP, with a very high accuracy.

To calculate mechanical properties of materials, such as elastic constants, the ideal strength, Young's modulus and Poisson's ratio, we developed a python general elastic calculation (PyGEC) package. For a 2D material, the stress-strain equation is obtained from the Hooke's law [equation (1)] under plane-stress condition[35].

$$\begin{bmatrix}\sigma_{xx}\\ \sigma_{yy}\\ \sigma_{xy}\end{bmatrix} = \begin{bmatrix}C_{11} & C_{12} & 0\\ C_{12} & C_{22} & 0\\ 0 & 0 & C_{66}\end{bmatrix}\begin{bmatrix}\varepsilon_{xx}\\ \varepsilon_{yy}\\ 2\varepsilon_{xy}\end{bmatrix} \quad (1)$$

We scanned the energy surface of materials in the strain range -1.5% $<\varepsilon_{xx}<$ 1.5%, -1.5% $<\varepsilon_{yy}<$ 1.5% and -1.0% $<\varepsilon_{xy}<$ 1.0%. The strain mesh grid was set to be 7×7×5. For 2D orthorhombic structures, the calculated four elastic stiffness constants $C_{11}$, $C_{12}$, $C_{22}$, $C_{66}$ (ESI, Table S1†), satisfy the necessary mechanical equilibrium conditions[36] for mechanical stability: $C_{11}C_{22}-C_{12}^2>0$ and $C_{11}$, $C_{22}$, $C_{66}$>0. Then, the orientation-dependent Young's modulus $E(\theta)$ and Poisson's ratio $\nu(\theta)$ are calculated as[3]:

$$\begin{cases} E(\theta) = \dfrac{Y_{zz}}{\cos^4\theta + d_2\cos^2\theta\sin^2\theta + d_3\sin^4\theta} \\ \nu(\theta) = \dfrac{\nu_{zz}\cos^4\theta - d_1\cos^2\theta\sin^2\theta + \nu_{zz}\sin^4\theta}{\cos^4\theta + d_2\cos^2\theta\sin^2\theta + d_3\sin^4\theta} \end{cases} \quad (2)$$

where $d_1$, $d_2$, $d_3$, $Y_{zz}$ and $\nu_{zz}$ are elastic constant related variables (ESI, Equation S1†). For bulk structure, we employ PyGEC to calculate elastic constants $C_{ij}$[37–39], and then use ElAM[40] code to analyze the Young's modulus and Poisson's ratio.

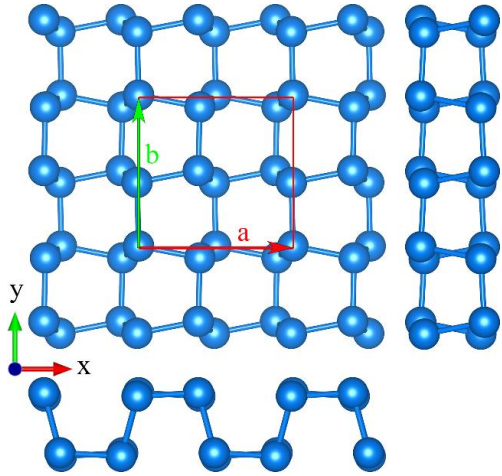

Fig. 1 The top and side views of the relaxed δ-P. The directions of two basis vectors *a* and *b* of the unit cell are

**RESULTS AND DISCUSSION**

The rectangular Wigner-Seitz cell of δ-P contains 8 atoms with Pmc2$_1$ symmetry. Our calculated lattice parameters for δ-P are a = 5.50 Å and b = 5.40 Å, consistent with previous theoretical study.[26] To explore the ideal tensile strength (the highest achievable stress of a defect-free crystal at 0 K) and critical strain (the strain at which ideal strength reaches)[41] of δ-P, an in-plane uniaxial tensile force is applied along either the *x* or the *y* direction. The stress-strain relationship for monolayer δ-P is presented in Fig. 2(a), where the tensile strain ranges from 0 to 30%. The ideal strengths are 7.08 GPa and 14.06 GPa in the *x* and *y* directions, respectively. The corresponding critical strains are almost of the same value (18.95%). We note that phonon instability may occur before mechanical failure. Such failure mechanism has been well studied in graphene where the phonon softening induced by Kohn anomaly occurs before the stress reaches its maximum.[42,43] To check the dynamic stability, we have calculated the phonon structure with external strains applied along x and y direction (See Fig. S1† in ESI), respectively. The results demonstrate that a strain up to 18.95% (equivalent to the mechanical fracture strength) along y direction is accessible whilst sustaining good phonon stability. As for x direction, when the strain reaches to 18.95%, a relative large imaginary frequency occurs. However, according to our calculations, δ-P can bear at least a strain of 15% along x direction.

Generally speaking, when a material is stretched in one direction, it tends to contract in the perpendicular direction. That is, when a tensile force is applied along one direction, the resulting vertical strain will be negative. This phenomenon is named the positive Poisson effect. However, for δ-P, we find that the tensile force [$F$, the insert of Fig. 2(b)] along *x* direction leads to a positive vertical strain along *y* direction, with the strain up to around 20%. So does the *y* direction. These results strongly indicate that δ-P has an anomalous auxetic effect with NPRs. What is more, the auxetic effect exists in a large strain range. In particular, this effect can sustain along y direction until the crack of the 2D material, since the critical strains (18.95%) is smaller than 20%.

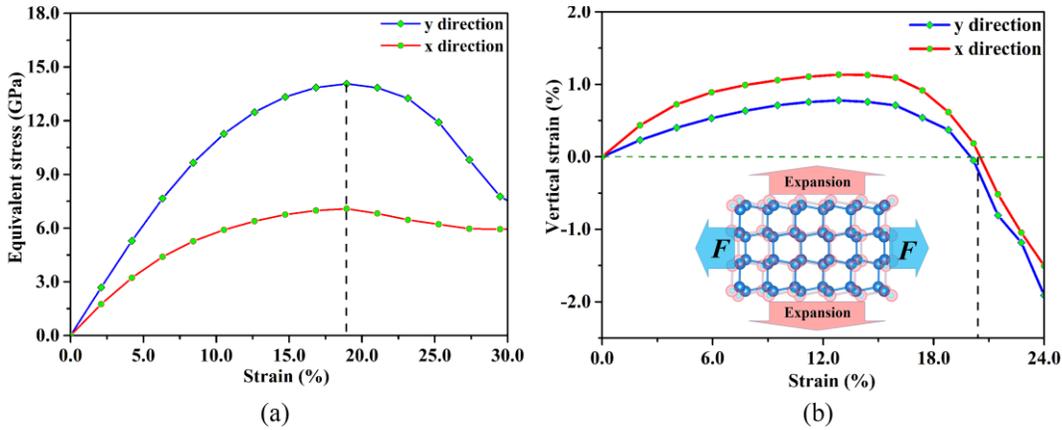

Fig. 2 (a) The strain-stress relation for monolayer δ-P. The strain is defined as $(r-r_0)/r_0$, where $r_0$ is the equilibrium lattice and r stands for strained one. Vertical dashed line indicates the critical strain and ideal strength. (b) The induced vertical strain in x direction when tensile strain

In the above section, we have gained a basic look into the mechanical response of δ-P by considering a uniaxial deformation only in the *x* or *y*-direction. However, due to the anisotropic geometric structure, the mechanical properties of δ-P are expected to be also highly orientation-dependent. In order to get a full understanding of mechanical properties of δ-P, the orientation-dependent Young's modulus and Poisson's ratio are calculated [Fig. 3(a) (b)]. Young's modulus represents the fully reversible stiffness response as in a linear elastic Hookean spring.[44] The 2D polar representation curve in Fig. 3(a) clearly indicates that the Young's modulus of δ-P is highly anisotropic, just as expected. The Young's modulus along the $y$ direction has a maximal value of 142.86 GPa and the minimum value of 62.28 GPa occurs along the direction indicated by the red arrow. For $x$ direction, it has a median value of 84.87 GPa. We notice that the maximal/minimal Young's modulus of δ-P and BP have the similar values, which may arise from their similar geometric structures. On the other hand, since P-P bond strength is weak and the compromised dihedral angles rather than the bond length stretch under a tensile force, the Young's modulus of both BP and δ-P are much smaller than those of graphene (1.0 TPa), $MoS_2$ (0.33 TPa) and BN (0.25 TPa),[2,45–47] which suggests that 2D δ-P is as flexible as BP.

From Fig. 3(b), one can see that the Poisson's ratios for δ-P are also spatially varying. In particular, δ-P exhibits a negative Poisson's ratio along the $x$, $y$ and their neighboring directions, while in the directions away from $x$, $y$, δ-P presents a normal positive Poisson's ratio. The maximal negative and positive Poisson's ratios are -0.267 and 0.29 along the $y$ direction and diagonal direction, respectively. As a comparison, we list the NPR values of recently reported 2D materials in Table 1. The NPR values of these 2D materials in the references were calculated by PBE functional, except for the bilayer-graphene's (VDW functional). We firstly recalculate the Poisson's ratio of all these selected 2D systems by using the PBE functional. The result shows that most of the Poisson's ratio values are negative and approximately equal to the reference ones, which further demonstrates the validity of our calculation. However, due to the special origin of NPR in bilayer-graphene,[13] the Poisson's ratio value calculated by PBE is positive (against the reference one), which proves to be wrong, since PBE gives a rather poor description of weak interlayer interaction in bilayer graphene. Considering that NPR values are possibly affected by the exchange and correlation functional forms, the VDW-DF functional[48,49] is employed to recalculate their Poisson's ratio values. As listed in Table 1, a NPR is obtained for bilayer-graphene. What's more, the variation trend of NPRs calculated with VDW-DF functional is similar to those with PBE. In this work, the NPR calculated by PBE functional will be adopted. According to the Table 1, it can be concluded that δ-P has the highest NPR value among these auxetic materials. It is also worth noting that the auxetic effect in BP[1] and bilayer-graphene[13] exists only along one direction, while δ-P holds for both the $x$ and $y$ directions.

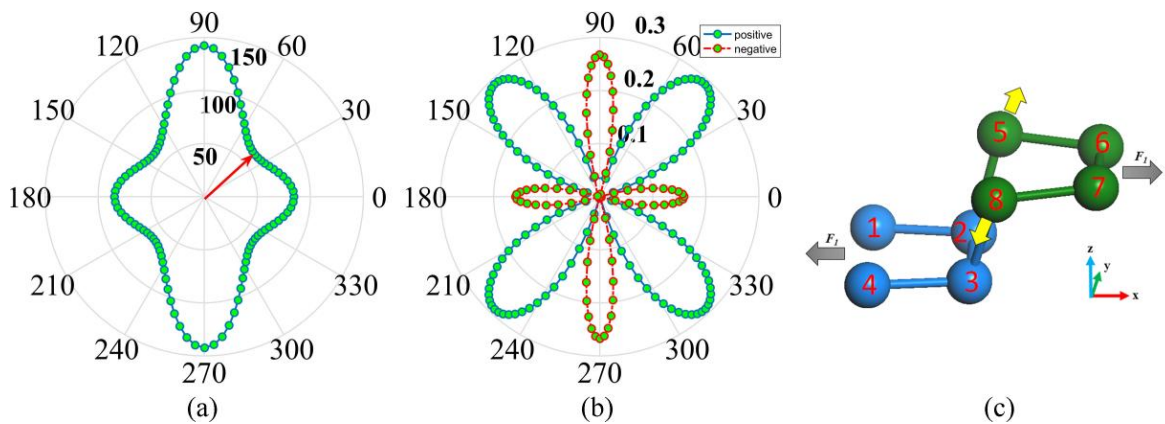

Fig. 3 Calculated orientation-dependent (a) Young's modulus E(θ) in GPa, (b) Poisson's ratio ν(θ). (c) The evolution of local structure of δ-P under a tensile force ($F_1$) along the x direction. δ-P expands along the y direction, indicated by the yellow arrows.

**Table 1** Comparison of negative Poisson's ratios ($v$) among recently predicted 2D materials with different functional.

| System | References | PBE | VDW-DF |
|---|---|---|---|
| δ-P | / | -0.267 | -0.400 |
| BP | -0.03(PBE[1]) | -0.048 | -0.130 |
| Penta-graphene | -0.07(PBE[24]) | -0.080 | -0.075 |
| Bilayer-graphene | -0.10(VDW[13]) | 0.642 | -0.198 |
| Borophene | -0.04(PBE[21]) | -0.040 | -0.032 |
| Penta-B$_2$N$_4$ | -0.02(PBE[22]) | -0.030 | -0.052 |
| SnSe | -0.17(PBE[19]) | -0.201 | -0.205 |

Due to the similarity of geometric structures between δ-P and BP, we may explain the origin of auxetic effect of δ-P in a similar way.[1] We plot the local skeleton structure of monolayer δ-P in Fig. 3(c), where the direction of tensile force and the motion of atoms are illustrated. Taking a tensile force $F_1$ in the $x$ direction as an example, the four top-most atoms 5, 6, 7 and 8 will move along the right arrow under the force. At the same time, the four bottom-most atoms 1, 2, 3 and 4 will move along the left arrow, which will lead to an increase in angles $\varphi_{125}$, $\varphi_{438}$, $\varphi_{387}$ and $\varphi_{652}$ and strain energy will be stored in these four angles. To accommodate the elongation in the $x$ direction, δ-P will contract along the $z$ direction. That is, atoms 1, 2, 3 and 4 will move upwards and atoms 5, 6, 7 and 8 will move downwards. For $\varepsilon_{xx}$=0.015, we find the thickness of monolayer δ-P is 2.214 Å, which is smaller than the initial value (2.238 Å). As a consequence, the 3-8 and 2-5 bonds become shorter, and part of strain energy will be transferred to these bonds. This is an unstable middle state, and a certain way is needed to release the strain energy. By carefully examining the original and final structures, we find that the angle $\varphi_{325}$ reduces by 0.206° while $\varphi_{832}$ increases by 0.585°. The net effect is that the distance between atom 5 and 8 becomes longer, which leads to the occurrence of an auxetic effect.

Although both δ-P and BP have honey comb like structure, the direction the auxetic effect holds in is different. To illustrate the subtle difference, we draw the wireframe sketches of δ-P and BP (See Fig. S2† in ESI). The side view indicates that the atoms in the top layer of δ-P have a distortion about 0.123 Å, so do the bottom ones. However, no distortion is observed in BP. Because of structure difference, the anisotropic ratio (defined by $E_y/E_x$) of BP (4.013)[3] is much larger than that of the δ-P's (1.683). In addition, the Poisson's ratio of BP along the groove direction is about 4 times larger than the perpendicular direction. As a consequence, a tensile force along the groove direction of BP will lead to a dramatic contraction on the perpendicular direction. To accommodate the contraction in this direction, the out-of-plane auxetic phenomenon occurs in BP, which has been demonstrated by experiment.[50] By contrast, the out-of-plane auxetic phenomenon of δ-P is suppressed by relatively small anisotropic ratio and appropriate in-plane auxetic effects.

Apart from BP, blue phosphorene has been synthesized on Au(111) surface,[51] however, theoretical study[52] predicted that it has no auxetic effect because there is no classic re-entrant structure existed. By investigating these 2D materials with NPR, it can be found that the auxetic effect is closely related to the atomic packing density and atom connectivity. Taking BP[1], SnSe[19] and δ-P as examples, we may find that all these 2D materials share the similarly hinge-like structure and this special structure is the necessary condition for their NPR. Thus, to synthesize other 2D phosphorene allotrope with NPR, a special attention should be paid to those with BP-like structures.

To investigate the possible quantum size effect on the mechanical properties of δ-P, we also calculate the Young's modulus and Poisson's ratios of double-, triple-, quadruple-layered and bulk δ-P. Firstly, we start with the relaxed monolayer δ-P structure, and explore their possible stacking orders (See Fig. S3† in ESI). The most favorable stacking configuration is found to be AB-stacking for double-layered δ-P, ABA-stacking for triple-layered δ-P, ABAB-stacking for quadruple-layered δ-P and AB-stacking for bulk δ-P. Then, the mechanical properties of multilayer and bulk δ-P with the most stable stacking configurations are calculated. As listed in Table 2, it can be summarized as follows: (1) the values of Young's modulus and Poisson's ratio are insensitive to the number of layers. This is because the interaction between δ-P layers is weak van der Waals interaction with no bonds formed. (2) The NPR along the *x* and *y* directions exists in all the cases, indicating the auxetic effect in δ-P is robust. (3) All structures have relatively small Young's modulus, with Ey much larger than Ex.

**Table 2** The calculated Young's modulus and Poisson's ratios for mono-, double-, triple-, quadruple-layered and bulk δ-P with PBE VDW-D2 functional. $v_{xy}$ is the Poisson's ratio with tensile strain applied in the *x* direction and the response strain in the *y* direction.

| Layers | Young's Modulus/GPa | | Poisson's ratio | |
|---|---|---|---|---|
| | $E_x$ | $E_y$ | $v_{xy}$ | $v_{yx}$ |
| Mono | 84.87 | 142.86 | -0.158 | -0.267 |
| Double | 89.83 | 150.37 | -0.125 | -0.209 |
| Triple | 89.55 | 148.49 | -0.128 | -0.213 |
| Quadruple | 90.55 | 148.32 | -0.129 | -0.211 |
| Bulk | 89.74 | 145.42 | -0.132 | -0.214 |

In addition, the electronic structures response to mechanical stress are also discussed in the Supplementary Information. The calculations show that a semiconductor-metal (M) transition can be observed by only changing the stress direction, and a direct semiconductor (D) to indirect semiconductor (I) transition can also be realized (ESI, Fig. S4 and S5†).

**SUMMARY**

In conclusion, first principles calculations have revealed that δ-P possesses highly anisotropic mechanical properties as well as fascinating characters of auxetic effects. The NPR value is higher than the previously reported 2D materials. The auxetic effect originating from the puckered structure exists in both x and y directions, and is insensitive to the number of stacked δ-P layers due to the weak van der Waals interaction. Besides, compared to other 2D materials, δ-P has a low Young's modulus and big critical crack strain, which indicates that 2D δ-P is also a material with good flexibility. These superior mechanical properties, along with the tunability of band gap, the direct-to-indirect semiconductor transition and semiconductor-to-metal transition, endow the δ-P a promising material for the design of nano-electromechanical devices, and we hope our study will stimulate further experimental effort in this subject.

## ACKNOWLEDGEMENT

This paper is financially supported by the National Key Research & Development Program of China (Grant No. 2016YFA0200604), by the National Natural Science Foundation of China (NSFC) (Grants No. 21421063, No. 21233007, and No. 21603205), by the Chinese Academy of Sciences (CAS) (Grant No. XDB01020300), by the Fundamental Research Funds for the Central Universities (Grant No. WK2060030023), by the China Postdoctoral Science Foundation (Grant No. BH2060000033). We used computational resources of Super-computing Center of University of Science and Technology of China, Supercomputing Center of Chinese Academy of Sciences, Tianjin and Shanghai Supercomputer Centers.

# Supplementary Information

δ-phosphorene: a two dimensional material with high negative Poisson's ratio

Haidi Wang,[a] Xingxing Li,[a,b] Pai Li[a] and Jinlong Yang[a,b*]

[a]*Hefei National Laboratory for Physical Sciences at the Microscale, University of Science and Technology of China, Hefei, Anhui 230026, China*

[b]*Synergetic Innovation Center of Quantum Information & Quantum Physics, University of Science and Technology of China, Hefei, Anhui 230026, China.*

*Email: jlyang@ustc.edu.cn*

Equation S1:

$$\begin{cases} v_{zz} = \dfrac{C_{12}}{C_{22}} \\ d_1 = \dfrac{C_{11}}{C_{22}} + 1 - \dfrac{C_{11}C_{22} - C_{12}^2}{C_{22}C_{66}} \\ d_2 = -(2\dfrac{C_{12}}{C_{22}} - \dfrac{C_{11}C_{22} - C_{12}^2}{C_{22}C_{66}}) \\ d_3 = \dfrac{C_{11}}{C_{22}} \\ Y_{zz} = \dfrac{C_{11}C_{22} - C_{12}^2}{C_{22}} \end{cases}$$

Equation S2:

The strained structures are obtained by the following equation.

$$\begin{cases} \varepsilon(\theta) = \sigma(\cos^4\theta + d_2\cos^2\theta\sin^2\theta + d_3\sin^4\theta) \\ \boldsymbol{\varepsilon}'(\boldsymbol{\theta}) = \begin{bmatrix} \cos\theta & -\sin\theta \\ \sin\theta & \cos\theta \end{bmatrix} \begin{bmatrix} 1+\varepsilon(\theta) & 0 \\ 0 & 1-\varepsilon(\theta)v(\theta) \end{bmatrix} \begin{bmatrix} \cos\theta & \sin\theta \\ -\sin\theta & \cos\theta \end{bmatrix} \\ \mathbf{R}' = \mathbf{R}\boldsymbol{\varepsilon}'(\boldsymbol{\theta}) \end{cases}$$

where $\sigma$ is the constant stress applied along the different directions ($\theta$). According to the $E(\theta)$, $v(\theta)$ and $\varepsilon(\theta)$, the strain matrix $\boldsymbol{\varepsilon}'(\boldsymbol{\theta})$ and lattice matrix $\mathbf{R}'$ can be calculated.

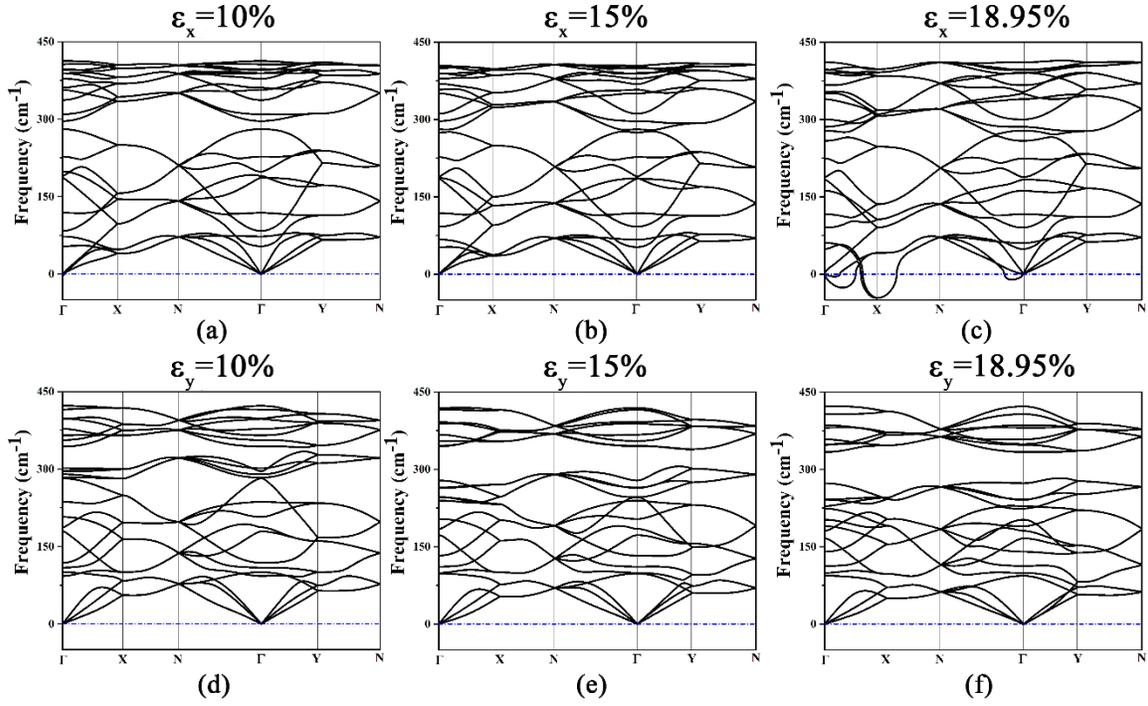

Fig. S1. Phonon bands of δ-P at different uniaxial strains.

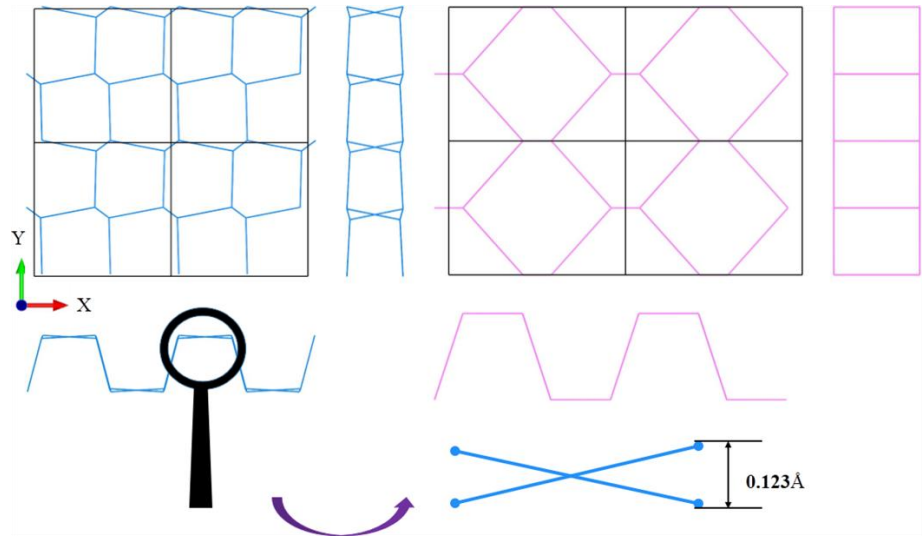

Fig. S2. The wireframe sketch of δ-P (right panel), BP (left panel), and locally enlarged structure δ-P (bottom panel).

A 2×2 supercell is adopted for the top and side view in Fig. S3. AB-stacking is the most favorable configuration for double-layered δ-P, being 16.7, 13.6 and 26.8 meV per atom lower than that of AA-, AB'-, and AC-stacking, respectively.

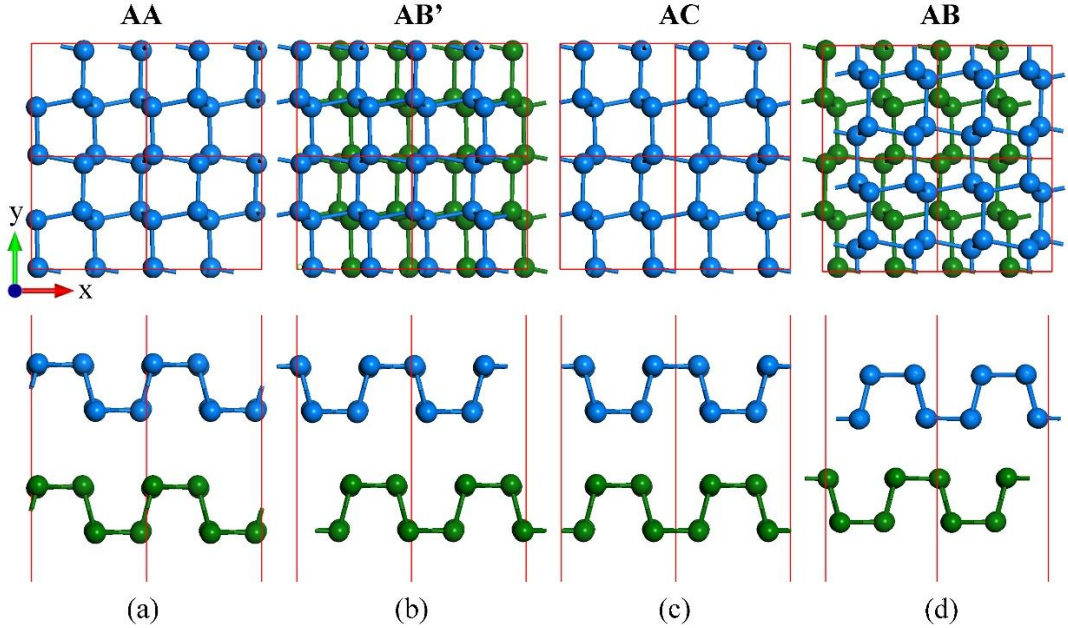

Fig. S3. Four stacking structures of double-layered δ-P. (a, b, c and d) Top views (upper panel) and side views (lower panel) of AA-, AB'-, AC-, AB-stacking, respectively.

The electronic structures response to mechanical stress are also discussed. We firstly explore the primitive δ-P. The band structure and associated density of states (DOS) is shown in Fig. S4. Consistent with previous study,[1] our results suggest that δ-P is a direct-gap semiconductor with a band gap of 0.50 eV, with the valence band maximum (VBM) and conduction band minimum (CBM) all located at Γ point. Meanwhile, both states are mainly contributed by $3p$ orbitals of P atoms. To better understand the electronic structure of δ-P, we also calculated the charge density of valence and conduction bands. As plotted in Fig. S4(b), a noticeable overlap can be found along the $x$ direction for valence band, while along the $y$ direction for conduction band respectively. This will result in different response of valance and conduction bands under external stress as we discussed later.

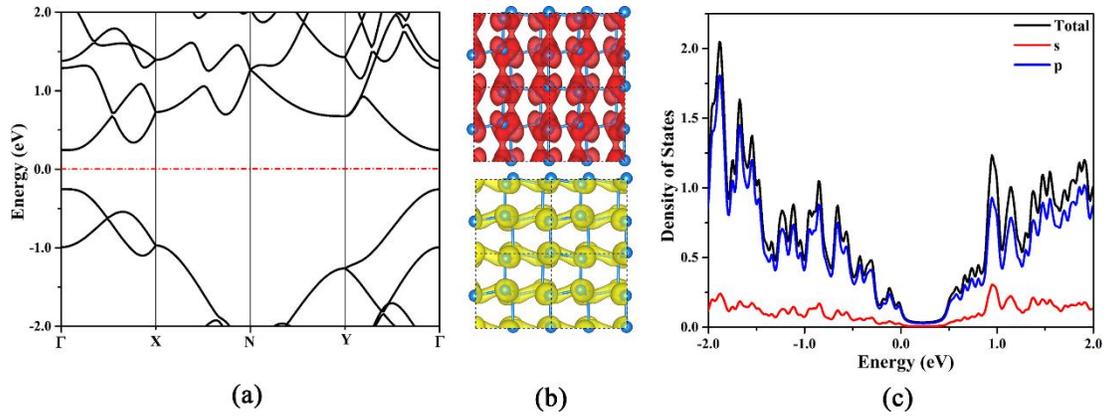

Fig. S4. (a) The band structure, dash-dotted line indicates the Fermi level. (b) The charge density corresponding to the valance band (yellow) and conduction band (red). (c) The total and projected DOS of primitive δ-P.

Then we turn to the discussion of electronic structures upon applying an external tensile stress. Under a small uniaxial stress of σ=1.0 GPa, the band gap of δ-P changes little. To adapt the different electronic device applications, a wide band gap engineering is desired. So we try to use a relatively large stress of σ=7.0 GPa. Such a stress is realizable, since a previous study[2] showed that a maximum stress of about 14 GPa is accessible for graphene. As is shown in Fig. S5(a), by gradually changing the stress direction, the VBM position moves up firstly when the stress direction angle is smaller than 40 °, then it declines, while the CBM presents an opposite change. To gain a full knowledge of band gap variation, we plot the band gap vs. angle relationship in Fig. S5(b). It's obvious that the band gap strongly depends on the applied stress direction, which further demonstrates the strong anisotropy of δ-P. In detail, the band gap ranges from 0.0 to 0.467 eV with a maximum band gap along the *y* direction and a zero band gap around θ=45°, becoming metallic. As a consequence, four band gap valleys are obtained. The calculations show that a semiconductor-metal (M) transition can be observed by only changing the stress direction, and a direct semiconductor (D) to indirect semiconductor (I) transition can also be realized. A schematic light disk in Fig. S5 (c) clearly illustrate a 'MID' loop formed when the stress direction scans from 0° to 360 °. It should be pointed out that these are DFT results, and the predictions are only qualitative.

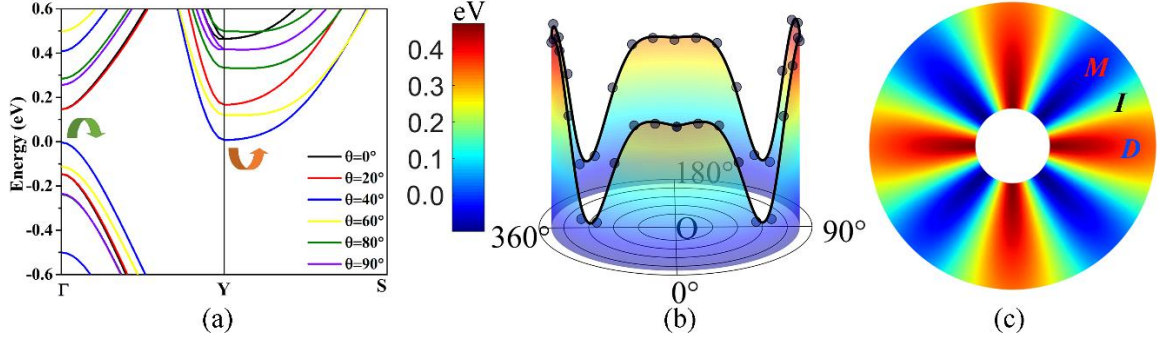

Fig. S5 (a) Band structure of δ-P under a stress of σ = 7.0 GPa in various directions. (b) Band gaps and spline fitting curve. (c) A color map of direct semiconductor (D) to indirect semiconductor (I) transition, and semiconductor to metal (M) transition tunable by changing the applied stress direction.

Table S1. The calculated elastic stiffness constants for mono-layered δ-P from PyGEC.

| unit | elastic stiffness constants | | | |
|---|---|---|---|---|
| | $C_{11}$ | $C_{22}$ | $C_{66}$ | $C_{12}$ |
| GPa | 88.64 | 149.21 | 24.50 | -23.71 |

In a 2D system, the stress calculated from the DFT has to be modified to avoid the force being averaged over the entire simulation cell including the vacuum slab[3]. In order to compare Young's modulus among different layered-structures, the length along z direction is rescaled by $\alpha\bar{h}$, where $\bar{h}$ is mean interlayer distance and α denotes the number of layer.

Table S2. The calculated lattice constants *a* and *b*, mean interlayer distance $\bar{h}$, and bonding energies Δ*E* of layered and bulk structure for δ-P.

| System | Stacking order | Lattice constants/Å | | $\bar{h}$/ Å | Δ*E*/meV/atom |
|---|---|---|---|---|---|
| | | a | b | | |
| Mono | A | 5.50 | 5.40 | / | 0.00 |
| Double | AB | 5.49 | 5.40 | 5.09 | -49.5 |
| Triple | ABA | 5.48 | 5.40 | 5.10 | -66.8 |
| Quadruple | ABAB | 5.47 | 5.40 | 5.19 | -75.4 |
| Bulk | AB | 5.46 | 5.39 | 5.11 | -101.4 |

Table S3. The calculated elastic stiffness constants for bulk δ-P.

| unit | elastic stiffness constants | | | | | | | | |
| --- | --- | --- | --- | --- | --- | --- | --- | --- | --- |
| | $C_{11}$ | $C_{22}$ | $C_{33}$ | $C_{44}$ | $C_{55}$ | $C_{66}$ | $C_{12}$ | $C_{13}$ | $C_{23}$ |
| GPa | 95.76 | 151.82 | 65.06 | 25.32 | 24.58 | 27.77 | -17.09 | 14.77 | 11.82 |